\documentclass[prb,amsmath,amssymp,showpacs, twocolumn, 
							showkeys, 
								]{revtex4}
\usepackage{bm} 
\usepackage{graphicx}
\usepackage{subfigure}
\usepackage{dcolumn}

\newcommand{\ri}{\mathrm{i}}
\newcommand{\rd}{\mathrm{d}}
\newcommand{\bo}{\hat{b}^{\phantom\dag}}
\newcommand{\ba}{\hat{b}^{\dag}}
\renewcommand{\ao}{\hat{a}^{\phantom\dag}}
\renewcommand{\aa}{\hat{a}^{\dag}}
\newcommand{\Ho}{\hat{H}}
\newcommand{\Vo}{\hat{V}}
\newcommand{\So}{\hat{S}}
\newcommand{\Ao}{\hat{A}}
\newcommand{\Bo}{\hat{B}}
\newcommand{\ho}{\hat{h}}
\newcommand{\vo}{\hat{v}}
\newcommand{\no}{\hat{n}}
\newcommand{\la}{\langle}
\newcommand{\ra}{\rangle}
\newcommand{\be}{\begin{equation}}
\newcommand{\ee}{\end{equation}}
\newcommand{\bes}{\begin{eqnarray}}
\newcommand{\ees}{\end{eqnarray}}
\newcommand{\tr}{\text{tr}}

\begin{document}

\title{Process chain approach to high-order perturbation calculus for quantum
lattice models}

\author{Andr\'{e} Eckardt}
\email[email: ]{andre.eckardt@icfo.es}
\affiliation{ICFO-Institut de Ci\`encies Fot\`oniques, 
Mediterranean Technology Park,
E-08860 Castelldefels (Barcelona), Spain}
\date{April 29th, 2009}

\begin{abstract}
A method based on Rayleigh-Schr\"odinger perturbation theory is developed that
allows to obtain high-order series expansions for ground-state properties of
quantum lattice models. The approach is capable of treating both lattice
geometries of large spatial dimensionalities $d$ and on-site degrees of freedom
with large state space dimensionalities. It has recently been used to accurately
compute the zero-temperature phase diagram of the Bose-Hubbard model on a
hypercubic lattice, up to arbitrary large filling and for $d=2$, 3 and greater 
[Teichmann {\it et al.}, Phys.\ Rev.\ B {\bf 79}, 100503(R) (2009)].
\end{abstract}
\maketitle
\section{Introduction}
Although rather simple, quantum lattice gas or spin models are known to give rise
to complex strongly correlated many-body physics. 
Currently the interest in these paradigmatic models of condensed matter physics
is stronlgy amplified by the fact that it just becomes experimentally feasable
to ``engineer'' many of them with ultracold atoms in optical lattice potentials
\cite{LewensteinEtAl07,BlochDalibardZwerger08}. 
Quantum lattice models are composed of small elementary sub-systems (sites), a
spin or the occupation of a single-particle state, arranged in a lattice. The
geometry of the lattice is reflected in the fact that only neighboring
sub-systems are coupled to each other directly. In general the Hamiltonian is of
the form
	\be\label{eq:Hgen}
	\Ho = \sum_i \ho_i 
		+\sum_{\la ij\ra}\vo_{ij},
	\ee
where the on-site term $\ho_i$ acts on the local state space of lattice site
$i$ only, while the coupling term $\vo_{ij}$ operates on both sites $i$ and
$j$ with the sum $\sum_{\la ij\ra}$ running over all pairs of neighboring
sites $\la ij\ra$ (bonds).

As the state-space dimensionality of the full system increases exponentially
with the number of sites $M$, generally the treatment of models
like~(\ref{eq:Hgen}) is a hard problem. However, sometimes there exist suitable
approximation schemes giving accurate results in certain regimes. One of them
is given by high-order series expansions, obtained by automation of the usual
Rayleigh-Schr\"odinger perturbation calculus. Such expansions have proven to be
a useful tool for the investigation of zero-temperature properties
\cite{GelfandSingh00,OitmaaHamerZheng}. In the simplest case, the coupling
terms $\vo_{ij}$ are considered as perturbation. Starting from a product
$|g\ra=\prod_i|n_i\ra\equiv|\{n_i\}\ra$ of local eigenstates $|n_i\ra$ with
$\ho_i|n_i\ra=\varepsilon_i|n_i\ra$, physical quantities like
expectation values or static susceptibilities (characterizing the state evolving
from $|g\ra$ adiabatically when the perturbation is switched on) can be expanded
in high-order power series with respect to a dimensionless coupling parameter.
Also quantum phase transitions \cite{Sachdev}, {\it i.e}.\ abrupt changes of the ground-state
structure, occurring when a certain parameter $\alpha$ passes a critical value
$\alpha_\text{c}$, can be inferred from such expansions.

A widely and successfully used algorithm for the application of perturbation
calculus to such lattice models is given by the connected cluster expansion
\cite{Marland81,GelfandEtAl90}: As a first step one has to determine all
sub-sets $C$ (connected clusters) of mutually connected bonds $\la ij\ra$
possessing not more than $\nu_\text{max}$ elements, with $\nu_\text{max}$ being
the largest order of the expansion to be considered. Moreover, all connected
sub-clusters (and sub-sub-clulsters, \ldots) of each of these connected 
clusters have to be identified. The second step consists of applying the
perturbation calculus iteratively to each of the small sub-systems
$\Ho_C= \sum_{i\in C'}\ho_i  +\sum_{\la ij\ra\in C}\vo_{ij}$ corresponding to
the connected clusters $C$, with $C'$ containing all sites connected to
the bonds of $C$. Finally, these results can be combined to
give the desired expansion.

However, when trying to apply the connected cluster formalism to the
Bose-Hubbard model \cite{FisherEtAl89}, describing bosonic particles with
short-range interaction moving in a lattice of single-particle orbitals
(sites), one encounteres two difficulties: (i)~Finding all connected clusters as
well as their sub-clusters is a demanding task for three-dimensional lattice
geometries (and even more for spatial dimensionalities $d>3$ that might be
interesting in order to probe the convergence to the mean-field limit).
(ii)~The dimensionalities $\mathcal{D}_C$ of the connected cluster
state spaces (that have to be considered after the particle number conservation
of the cluster Hamiltonians $\Ho_C$ has already been taken into account) grow
extremely fast with respect to the averaged lattice site occupation
(filling)~$n$. As a consequence, an iterative evaluation of the perturbation
series up to 8th (10th) order is hindered already for the moderate filling
$n=5$ ($n=3$) by the fact that it requires a representation of the cluster
state spaces on the machine
\footnote{In order to compute corrections to the ground state energy in $\nu$th
order, one has to consider connected clusters containing up to $M_C=\nu$ sites,
such that $\mathcal{D_C}$ is given by the number of possibilities to place
$N_C=nM_C$ indistinguishable particles on these sites. 
If about 500 bytes are needed per state space dimension, this gives 100 gigabyte
of required RAM already for the filling of either $n=3$ in 10th order or $n=5$
in 8th order.}.
While difficulty (i) appears for every quantum lattice model of large spatial
dimensionality $d\ge3$, difficulty (ii) is noticed for systems with the
relevant local on-site state-space dimensionalities being too large.
In the following, a method based on Kato's formulation of Rayleigh-Schr\"odinger
perturbation calculus \cite{Kato49} will be described that circumvents both
difficulties (i) and (ii). Here, perturbative corrections are obtained as sums
over chains of processes acting in the ``classical'' space containing the
unperturbed states but not their superpositions. These process chains, in turn,
are generated from paths through the $d$-dimensional lattice. As an example,
using this approach one is able to accurately compute the Bose-Hubbard phase
diagram at any integer filling $n$ and for spatial dimensionalities $d=2$, 3,
and larger \cite{TeichmannEtAl09}. An implementation of the method presented
here is straightforward and can be accomplished from scratch with reasonable
effort. The present approach is found to be related to the ones
described previously in Refs.~\onlinecite{ParrinelloEtAl73,HamerIrving84,
PrelovsekEtAl90} where contributions to the perturbation series are equally
expressed in terms of sequences of processes. Differences to these approaches
lie both in the way contributions are organized\footnote{By taking Kato's
expression (\ref{eq:Kato}) as a starting point here subcluster subtractions
(or cumulant corrections) do not have to be performed on the level of each
single diagram, as it is the case in Refs.~\onlinecite{ParrinelloEtAl73} and
\onlinecite{HamerIrving84}. In contrast, the approach briefly sketched in
Ref.~\onlinecite{PrelovsekEtAl90} is more similar to the one presented here; it
does, however, not include the (huge) reduction of terms contributing to the
general perturbation expansion that will be described in subsection
\ref{sec:reduction}.} and in the generation of diagrams from paths through the
lattice.

For clarity, the process chain method will be introduced in terms of the
simple Bose-Hubbard model \cite{FisherEtAl89}. The generalization of the
method to more involved or just different lattice Hamiltonians (like those of
Heisenberg-type spin models) is, however, straightforward; later on the
properties of a general model amenable to the approach will be sketched. The
Bose-Hubbard Hamiltonian reads
	\bes\label{eq:BH}
	\Ho_\text{BH} &=&\sum_i \left[\frac{U}{2} 
	\no_i(\no_i-1)+(\varepsilon_i-\mu)\no_i\right]
	\nonumber\\&&
			-\,J\sum_{\la ij\ra}\left(\ba_i\bo_j
				+\ba_j\bo_i \right),
	\ees
with $\ba_i$, $\bo_i$, and $\no_i\equiv\ba_i\bo_i$ denoting the bosonic
creation, annihilation, and number operator for a single-particle orbital
located at site $i$. The first line of Eq.~(\ref{eq:BH}) includes an on-site
interaction characterized by the energy cost~$U$ for each pair of particles
occupying the same site. Moreover, it assigns the local potential energy
$\varepsilon_i-\mu\equiv-\mu_i$ to particles sitting at site $i$, including an
overall chemical potential $\mu$ introduced to control the total particle
number $N$. The terms of the second line implement the kinetics, being exhausted
by tunneling of particles between neighboring sites. Although rather simple,
this model provides a quantitative description of ultracold bosonic alkali
atoms in optical lattice potentials~\cite{JakschEtAl98,BlochDalibardZwerger08}.
It shows quantum phase transitions between a gapless, compressible superfluid
phase with (quasi) long-range order present at large values of the ratio $J/U$
and various gapped incompressible Mott-insulator phases (at sufficiently small
$J/U$) with exponentially decaying correlations, each characterized by an
integer filling factor $n=\sum_i\la\no_i\ra/M$ (depending on the chemical
potential $\mu/U$)\cite{FisherEtAl89,Sachdev}.

This paper is organized as follows: Section \ref{sec:perturbation} is devoted
to the general perturbation expansion. It is briefly reviewed how expectation
values and static susceptibilities can be obtained via the computation of energy
corrections. Starting from Kato's formulation of the $\nu$th energy
correction\cite{Kato49}, then a way to reduce the number of terms that have to
be taken into account to a minimum is described. This is the first step that
has to be accomplished on the machine. Finally, the perturbative corrections are
written as sums over process chains. This formulation will serve as a 
fruitful starting point for customizing perturbation calculus to quantum
lattice models in the way developed in section \ref{sec:lattice}. The main
conceptual step of section \ref{sec:lattice} consists in considering groups of
operations (to be visualized by diagrams) such that each sequence of the
operations contained in a group/diagram gives a process chain contributing to
the desired perturbative correction. It is explained that the generation of the
relevant diagrams can be put down to the rather simple task of generating paths
through the lattice (this allows one to address also large $d$), and it is
described how the evaluation of diagrams can be performed. These are the two
final steps to be implemented on the computer. In order to give a comprehensive
presentation of the approach, the computation of perturbative corrections is
first discussed in the context of the Bose-Hubbard model (\ref{eq:BH}).
Following along the lines given by this instructive example the application of
the method to the large class of quantum lattice models described at the end of
section \ref{sec:lattice} should be straightforward. Section \ref{sec:summary} 
summarizes the basic steps of the approach, before section \ref{sec:conclusion}
closes with concluding remarks.

\section{\label{sec:perturbation} General perturbation expansion}
\subsection{Problem}
Consider a system described by the time-independent Hamiltonian
	\be
	\Ho = \Ho_0 + \lambda\Vo
	\ee	
consisting of an unperturbed part $\Ho_0$ that is already diagonalized,
	\be
	\Ho_0 = \sum_e E_e |e\ra\la e|,
	\ee
and a perturbation
	\be\label{eq:V}
	\Vo = \sum_{e,e'} V_{e',e} |e'\ra\la e|
	\ee 
multiplied by a dimensionless factor $\lambda$ finally to be set equal to 1.
The standard non-degenerate Rayleigh-Schr\"odinger perturbation calculus
\cite{Messiah2} gives a power law expansion 
	\be\label{eq:EE}
	E_G 
	  	= E_g + \lambda  E_g^{(1)} + \lambda^2  E_g^{(2)} + \cdots.
	\ee
for the energy of the state $|G\ra$ evolving adiabatically from an
eigenstate~$|g\ra$ of the unperturbed Hamiltonian ({\it e.g}.~its ground state)
when the perturbation is switched on. In the following the question of
convergence will not be discussed, but an algorithm for the computation of such
series expansions for many-body quantum lattice models is deviced.

Usually, one is not interested in high-order perturbative corrections to the
\emph{state} of a many-body system, containing a vast amount of mostly unwanted
information. One rather wishes to compute selected quantities like expectation
values $\la G|\Ao|G\ra$. It is known, however, that the computation of ground
state expectation values and static susceptibilities can be reduced to the
evaluation of energy corrections. Introducing the extended Hamiltonian
	\be
	\Ho_{AB} = \Ho_0 + \lambda\Vo + x\Ao + y\Bo,
	\ee
with operators $\Ao$ and $\Bo$, a perturbative treatment of
$\Vo'\equiv\lambda\Vo + x\Ao + y\Bo$ yields the expansion
	\be
	E_{G_{AB}} 
		= \sum_{\nu,m,k} E_g^{(\nu,m,k)}  \lambda^\nu x^m y^k
	\ee
for the energy of the perturbed eigenstate $|G_{AB}\ra$ evolving
from $|g\ra$. [In the case of non-hermitian operators $\Ao$ one
can consider instead the hermitian ones
$\Ao^{(+)}\equiv\frac{1}{2}(\Ao+\Ao^\dag)$ and
$\Ao^{(-)}\equiv\frac{1}{2\ri}(\Ao-\Ao^\dag)$ such that
$\Ao=\Ao^{(+)}+\ri\Ao^{(-)}$.] The low-order coefficients with respect to $x$
and $y$ then give series in powers of $\lambda$ for the expectation value
	\be\label{eq:ExpA}
	\la G|\Ao|G\ra 
	=\sum_\nu \lambda^\nu E_g^{(\nu,1,0)},
	\ee
and the static susceptibilities
	\be\label{eq:SusAB}
	\chi_{AB} 
		=\sum_\nu \lambda^\nu E_g^{(\nu,1,1)}
	\ee
and
	\be
	\chi_{AA} 
			= \frac{1}{2}\sum_\nu \lambda^\nu E_g^{(\nu,2,0)}
	\ee
describing the linear response of $\la\Ao\ra$, 
when the full Hamiltonian $\Ho$ is perturbed by $\Bo$ or $\Ao$, respectively.
One can obtain these relations, {\it e.g}., by using the Hellmann-Feynman
theorem $\frac{\rd}{\rd z}\la\psi(z)|\Ho(z)|\psi(z)\ra
=\la \psi(z)|[\frac{\rd}{\rd z}\Ho(z)]|\psi(z)\ra$ [valid if $|\psi(z)\ra$ is a
normalized eigenstate of $\Ho(z)$] or by interpreting them as first order
energy corrections to the unperturbed problem given by the full
Hamiltonian $\Ho$. Hence, solely by using a formalism for the evaluation of
energy corrections, one can obtain series expansions for many important
quantities characterizing the system.

\subsection{Minimal expression for the $\nu$th order energy correction
\label{sec:reduction}}
The $\nu$th energy correction appearing in Eq.~(\ref{eq:EE}) is given by Kato's
closed expression \cite{Kato49} (see also Ref.~\onlinecite{Messiah2})
	\be\label{eq:Kato}
	E_g^{(\nu)} = \sum_{(\nu-1)}\tr\left\{\So^{\alpha_{\nu+1}}\Vo\So^{\alpha_{\nu}}
		 \cdots\Vo\So^{\alpha_3}\Vo\So^{\alpha_2}\Vo\So^{\alpha_{1}} \right\}
	\ee
with the sum $\sum_{(\nu-1)}$ running over all combinations of the $\nu+1$
non-negative integers $\alpha_k$ such that $\sum_{k=1}^{\nu+1}\alpha_k=\nu-1$, 
$\tr\{\cdot\}\equiv \sum_e\la e|\cdot|e\ra$, and
	\be\label{eq:S}
	\So^\alpha \equiv \left\{   \begin{array}{ll}				
			-|g\ra\la g|&	\quad\text{for}\quad \alpha = 0 \\
			\sum_{e\ne g}\frac{|e\ra\la e|}
					{\left(E_g-E_{e}\right)^\alpha}
			& \quad\text{for}\quad \alpha\ge1	\;.	
					\end{array}   \right.
	\ee
Since on the r.h.s.\ of Eq.~(\ref{eq:Kato}) there are always at least two 
$\alpha_i$ equal to zero, by cyclic permutation under the trace and using
	\be
	\So^\alpha\So^{\alpha'}= \left\{   \begin{array}{ll}
		-\So^0  &\quad\text{for}\quad \alpha=0\text{ and }\alpha'=0\\
		0	&\quad\text{for}\quad \alpha=0\text{ and }\alpha'\ne0\\
			&\quad\quad\text{or } \alpha\ne0\text{ and }\alpha'=0 \\
		\So^{\alpha+\alpha'}  
			&\quad\text{for}\quad \alpha\ne0\text{ and }\alpha'\ne0\\	
					\end{array}   \right.
	\ee
the energy correction can always be expressed as a sum over expectation values
with respect to the unperturbed state $|g\ra$. This gives
	\be\label{eq:Expect}
	E_g^{(\nu)} = \sum_{(\nu-1)}G_{\{\alpha_k\}} \la g|\Vo \So^{\alpha_{\nu-1}}\Vo
		 \cdots\So^{\alpha_2}\Vo\So^{\alpha_{1}}\Vo|g\ra
	\ee
with $\sum_{k=1}^{\nu-1}\alpha_k=\nu-1$ according to the constraint of the
sum (\ref{eq:Kato}) and with (not uniquely determined) weight factors
$G_{\{\alpha_k\}}$ taking into account how often each matrix element is
generated with positive and negative prefactor during the elimination of the
trace. Below the short hand
	\be\label{eq:sh}
	(\alpha_\ell,\ldots,\alpha_2,\alpha_1)\equiv	
	\la g|\Vo \So^{\alpha_\ell}\Vo\cdots
		 \So^{\alpha_2}\Vo\So^{\alpha_1}\Vo|g\ra,
	\ee
for the matrix elements appearing in the sum (\ref{eq:Expect}) will be used,
with $()\equiv\la g|\Vo|g\ra$ denoting the first order energy correction.

An example for an expression like (\ref{eq:Expect}) is given by the formula
\cite{Bloch58} (see also Ref.~\onlinecite{Messiah2})
	\be\label{eq:Bloch}
	E_g^{(\nu)} = {\sum_{(\nu-1)}}' (\alpha_{\nu-1},\ldots,\alpha_2,\alpha_1),
	\ee
where all $G_{\{\alpha_k\}}$ are either zero or one, as it is encoded in the
set of constraints $\sum_{k=1}^s\alpha_k\ge s$ with $s=1,2,\ldots(\nu-2)$,
additional to the requirement $\sum_{k=1}^{\nu-1}\alpha_k\ge \nu-1$.
A further way to obtain an expression of type (\ref{eq:Expect}) similar to
formula (\ref{eq:Bloch}) is to start with the matrix element $(1,1,\ldots,1,1)$,
with all $\alpha_k=1$, and to generate successively further matrix elements to
be considered by applying the recursive scheme described in
Ref.~\onlinecite{Huby61}.

Many matrix elements $(\alpha_{\nu-1},\ldots,\alpha_2,\alpha_1)$ that appear in
the sums (\ref{eq:Expect}) or (\ref{eq:Bloch}) give identical contributions:
Writing explicitly $\So^0=-|g\ra\la g|$, each matrix element
$(\alpha_{\nu-1},\ldots,\alpha_2,\alpha_1)$ breaks up into elementary matrix
elements (EME) $(\beta_\ell,\ldots,\beta_2,\beta_1)$, containing strictly
positive ({\it i.e}.\ non-zero) integers $\beta_i$ only. Thus, {\it e.g}., one
has
	\bes
	&(1,1,0,0,3,0,2,1)&=-(1,1)()(3)(2,1)
	 \\\nonumber
	 = &(1,1,0,0,3,0,1,2)&=-(1,1)()(3)(1,2)
	  \\\nonumber
	 = &(1,1,0,0,2,1,0,3)&=-(1,1)()(2,1)(3)
	 \\\nonumber
	 = &(1,1,0,0,1,2,0,3)&=-(1,1)()(1,2)(3)
 	 \\\nonumber
	 =&\cdots& .
	\ees 
Here two basic operations were applied leaving the expression unchanged, the
permutation of EMEs and the ``reflection'' of an EME,
$(\beta_\ell,\ldots,\beta_2,\beta_1)\to(\beta_1,\beta_2,\ldots,\beta_\ell)$.
The latter is allowed since both $\Vo$ and the $\So^\alpha$ are hermitian. 
Hence, one has families of matrix elements
$(\alpha_{\nu-1},\ldots,\alpha_2,\alpha_{1})$ with all members giving the same
contribution. 
In contrast, different EMEs $(\beta_\ell,\ldots, \beta_2, \beta_1)$ generally
give different contributions, if, by convention, one considers EMEs differing
just by a ``reflection'' to be identical. Accordingly, each family of equally
contributing matrix elements $(\alpha_{\nu-1},\ldots,\alpha_2,\alpha_{1})$ is
uniquely determined by a set of EMEs. In order to take into account only one
representative of each family the energy correction can be rewritten as
	\be\label{eq:RedEn}
	E_g^{(\nu)} = \sum_{(\nu-1)} G_{\{\alpha_k\}}^\text{min} 
				(\alpha_{\nu-1},\ldots,\alpha_2,\alpha_{1})
	\ee
with a minimum number of non-vanishing weight factors
$G_{\{\alpha_k\}}^\text{min}$. Obtaining such a minimal expression for the
orders to be considered is the first problem that has to be solved on a
computer for an implementation of the method described here. Since so far one is
dealing with the general perturbation expansion, this step has to be performed
once only. 

\begin{table}[t]
\centering
\begin{tabular}{|r|r|r|}
\hline
order $\nu$ 
	& $K_\nu$ (\# of terms)  
		&$K'_\nu$ (\# of restrict. terms)  \\\hline\hline
1	&1	&0	\\\hline
2	&1	&1	\\\hline
3	&2	&1	\\\hline
4	&4	&2	\\\hline
5	&10	&3	\\\hline
6	&22	&7	\\\hline
7	&53	&12	\\\hline
8	&119	&26	\\\hline
9	&278	&47	\\\hline
10	&627	&97	\\\hline
11	&1,433	&180	\\\hline
12	&3,216	&357	\\\hline
13	&7,253	&668	\\\hline
14	&16,169	&1,297	\\\hline
15	&36,062	&2,427	\\\hline
16	&79,876	&4,628	\\\hline
17	&176,668&8,637	\\\hline
18	&388,910&16,260	\\\hline
19	&854,493&30,188	\\\hline
20	&---	&56,252	\\\hline
21	&---	&103,848\\\hline
22	&---	&191,873\\\hline
23	&---	&352,204\\\hline
24	&---	&646,061\\\hline
\end{tabular}
\caption{\label{tab:NumTerm} Minimum number $K_\nu$ of matrix elements that
have to be considered for the $\nu$th order energy correction. The number of
contributing terms is significantly reduced to $K'_\nu$ if the first-order
energy correction vanishes, $\la g|\Vo|g\ra=()=0$. The data for $\nu\ge12$ is
taken from Ref.~\onlinecite{Steenhoff07}.}
\end{table}

A routine (R1) for the generation of a minimal set of matrix elements
contributing to the sum~(\ref{eq:RedEn}) and their weights can be based on one
of two alternative approaches. The first one is to generate all matrix elements
appearing in an expression like (\ref{eq:Expect}) by starting either from
formula (\ref{eq:Kato}) or (\ref{eq:Bloch}), and then to identify members of
the same matrix element family by decomposition into EMEs. The second approach
is to generate all EMEs and all combinations of them describing families of
matrix elements $(\alpha_{\nu-1},\ldots,\alpha_2,\alpha_{1})$ appearing in the
sum (\ref{eq:Expect}). The weight factor $G_{\{\alpha_k\}}^\text{min}$ of a
given family can then be obtained from a simple expression
\cite{Steenhoff07}. The minimum number $K_\nu$
of matrix elements to be considered in $\nu$th order is listed in table
\ref{tab:NumTerm}. It is drastically reduced, if it is known that the first
order energy correction $\la g|\Vo|g\ra=()$ vanishes. Generally, given
knowledge about the vanishing of corrections in certain orders $\nu$ that are,
say, even, odd or smaller than a value $\nu_0$ can be used to reduce the number
of matrix elements that has to be taken into account. For example, if all
corrections appearing in orders smaller than $\nu_0$ are known to vanish, then
there is at most one relevant matrix element appearing in order $\nu_0$.

\subsection{Energy corrections as sums over process chains}
One can interpret each matrix element
$(\alpha_{\nu-1},\ldots,\alpha_2,\alpha_{1})$ appearing in Eq.~(\ref{eq:RedEn})
as a weighted sum $\sum_{\{e_i\}}$ over paths
$|g\ra\to |e_1\ra\to |e_2\ra\to\cdots\to |e_{\nu-1}\ra\to |g\ra$ in a
``classical'' space containing the unperturbed states $|e\ra$ but not their
superpositions. All paths lead from $|g\ra$ back to $|g\ra$ via $\nu-1$
intermediate states $|e_k\ra$: By plugging definitions~(\ref{eq:V}) and
(\ref{eq:S}) into Eq.~(\ref{eq:sh}), one obtains 
	\bes\label{eq:paths}
	(\alpha_{\nu-1},\ldots,\alpha_2,\alpha_{1})
	&=& \sum_{\{e_k\}} V_{g,e_{\nu-1}}W^{(\alpha_{\nu-1})}_{e_{\nu-1}}
	\cdots
	\\\nonumber&&\times\cdots	
	W^{(\alpha_2)}_{e_2}V_{e_2,e_1}W^{(\alpha_1)}_{e_1} V_{e_1,g}
	\ees
with 	
	\be\label{eq:EnergyWeights}
	W^{(\alpha)}_{e} \equiv 
		-\delta_{\alpha,0}\delta_{e,g}
		+(1-\delta_{\alpha,0})(1-\delta_{e,g})[E_g-E_{e}]^{-\alpha}.
	\ee
In that way a formulation involving huge quantum mechanical state spaces is
avoided. 

However, usually we have to deal with several perturbing terms at once,
$\Vo=\Vo^{(1)}+\Vo^{(2)}+\cdots$ with 
$\Vo^{(m)}=\sum_{e',e}V^{(m)}_{e',e}|e'\ra\la e|$, and we wish to keep track of
them independently. Therefore, it is convenient to reformulate 
Eq.~(\ref{eq:paths}) once more, namely as a sum over \emph{process chains} $P$
each of them being given by an ordered sequence
$V^{(m_1)}_{e_1,g}|e_1\ra\la g|$, $V^{(m_2)}_{e_2,e_1}|e_2\ra\la e_1|$, \ldots,
$V^{(m_{\nu})}_{g,e_{\nu-1}}|g\ra\la e_{\nu-1}|$ of basic processes
$V^{(m)}_{e',e}|e'\ra\la e|$ leading from $|g\ra$ back to $|g\ra$ in the
classical space of unperturbed states introduced above. One has
	\bes\label{eq:pro}
	(\alpha_{\nu-1},\ldots,\alpha_2,\alpha_{1})
	&=& \sum_P V^{(m_{\nu})}_{g,e_{\nu-1}}W^{(\alpha_{\nu-1})}_{e_{\nu-1}}
		\cdots
	\\\nonumber&&\times\cdots	
	W^{(\alpha_2)}_{e_2}V^{(m_2)}_{e_2,e_1}W^{(\alpha_1)}_{e_1}
							 V^{(m_1)}_{e_1,g},
	\ees
or, in combination with Eq.~(\ref{eq:RedEn}),
	\be\label{eq:EnPro}
	E_g^{(\nu)} =  \sum_P\sum_{\{\alpha_k\}} G_{\{\alpha_k\}}^\text{min}
		 V^{(m_{\nu})}_{g,e_{\nu-1}} 
		 	\cdots W^{(\alpha_2)}_{e_2}V^{(m_2)}_{e_2,e_1} 
				W^{(\alpha_1)}_{e_1}V^{(m_1)}_{e_1,g}
	\ee
for the $\nu$th order energy correction. 

The strategy for the computation of energy corrections proposed here is to
generate all process chains $P$ appearing in a given order $n$ in a first step,
and then to compute the contributions arising from each chain according to the
different sets $\{\alpha_k\}$ possessing non-vanishing weight factors
$G_{\{\alpha_k\}}^\text{min}$ in the general perturbation expansion in the
form given by Eq.~(\ref{eq:RedEn}). In the next section it is shown how both
steps can be accomplished efficiently for a lattice system with short-range
coupling.

\section{\label{sec:lattice}Lattice System}

\subsection{Bose-Hubbard problem}
In this section, the above formalism is applied to the Bose-Hubbard
Hamiltonian~(\ref{eq:BH}) on a hypercubic lattice geometry. All terms that are
diagonal with respect to the lattice-site occupation numbers
$n_i$ will be considered as unperturbed problem
	\be
	\Ho_0=\sum_i \left[\frac{1}{2}(\no_i-1)-\frac{\mu_i}{U}\right] \no_i,
	\ee 
and the remaining tunneling terms as perturbation
	\be
	\Vo=-\frac{J}{U}\sum_{\la ij\ra}\left(\ba_i\bo_j+\ba_j\bo_i\right).
	\ee 
Energies have been expressed in units of the positive interaction parameter $U$,
such that $J/U$ is identified to be the dimensionless coupling parameter. 
A basis of unperturbed states $|e\ra$ is given by the lattice-site
occupation-number states
	\be
	|\{n_i\}\ra \equiv \prod_i\frac{(\ba_i)^{n_i}}{\sqrt{n_i!}}
							|\text{vacuum}\ra.
	\ee
Let us assume that the state we want to investigate is the one evolving
adiabatically from the unperturbed state 
	\be
	|g\ra \equiv|\{n_i=g_i\}\ra 
	\ee 
when the perturbation is switched on. If $|g\ra$ denotes the unperturbed ground
state its occupation numbers $g_i$ minimize $[(g_i-1)/2-\mu_i/U] g_i$ and read
	\be
	g_i = \left\{   \begin{array}{ll}	
			0& 	\quad\text{if}\quad \mu_i/U<0 \\
			h&	\quad\text{if}\quad (h-1)<\mu_i/U<h \\
			(h-1)\text{ or }h& \quad\text{if}\quad\mu_i/U=h-1\ge0\;,
				\end{array}   \right.
	\ee
with non-negative integers~$h$. As long as the marginal case of integer
$\mu_i/U$ is avoided, this state is protected by an energy gap. 

In the following first a thorough discussion of the computation of
(actual) energy corrections is given as an instructive example. Then it will be
shown that this example already contains everything one needs also for the
computation of other quantities of interest, like single particle correlations
$\la\ba_i\bo_j\ra$, number correlations $\la\no_i\no_j\ra$, or the static
susceptibility $\chi_{\ao_0,\aa_0}$ for the annihilation and creation operators
of the condensate mode, $\ao_0$ and $\aa_0$. The divergence of the latter
indicates the quantum phase transition from a Mott-insulator to a superfluid 
\cite{NegeleOrland,dosSantosPelster09,TeichmannEtAl09}.
The approach can easily be generalized to more complicated or just different
lattice models. A general model that is amenable to the procedure described
below is sketched at the end of this section.

\subsection{Corrections to the energy}

\begin{figure}[t]\centering
\includegraphics[width = 0.3\linewidth]{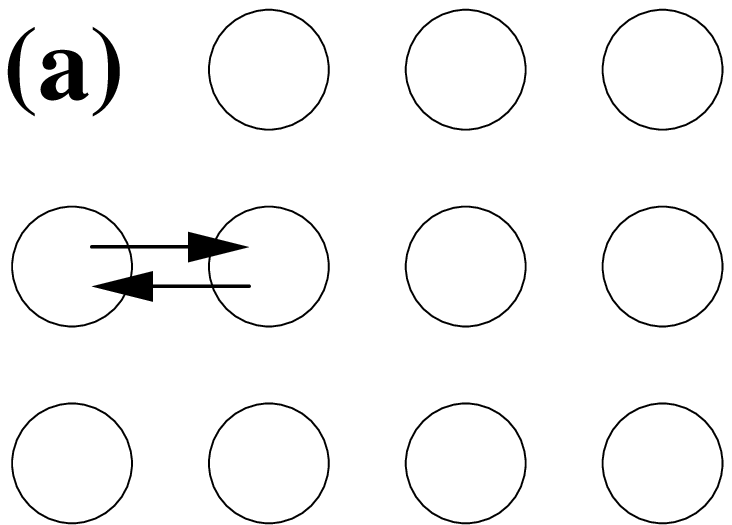}
\hspace{5ex}
\includegraphics[width = 0.3\linewidth]{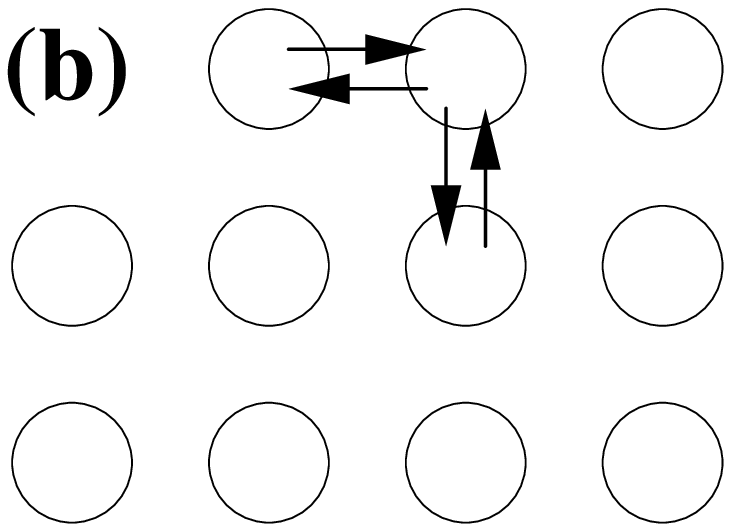}
\\\vspace{5ex}
\includegraphics[width = 0.3\linewidth]{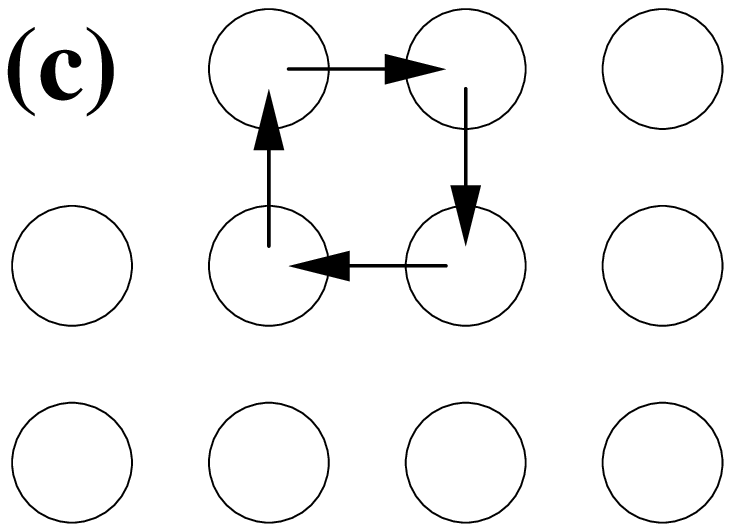}
\hspace{5ex}
\includegraphics[width = 0.3\linewidth]{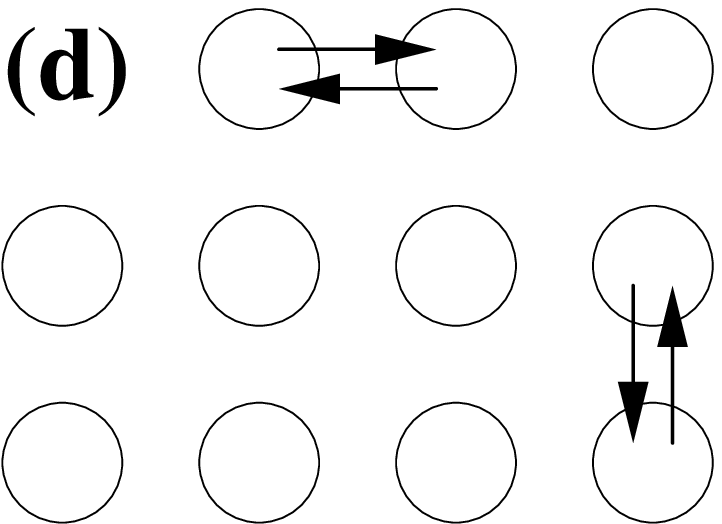}
\\\vspace{5ex}
\includegraphics[width = 0.3\linewidth]{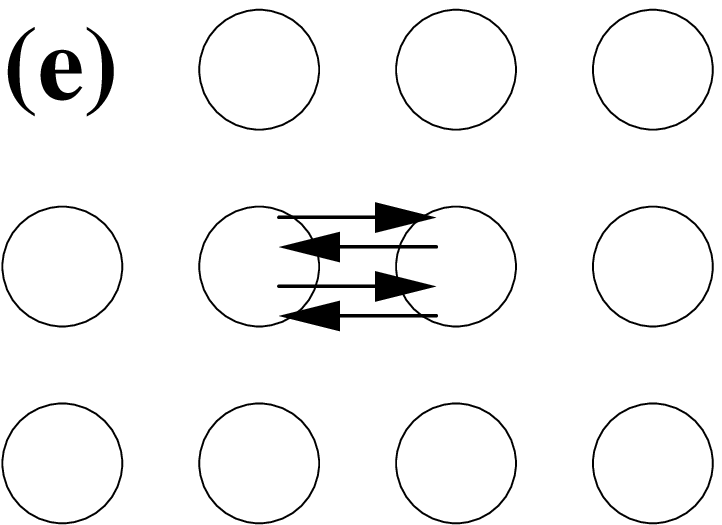}
\hspace{5ex}
\includegraphics[width = 0.3\linewidth]{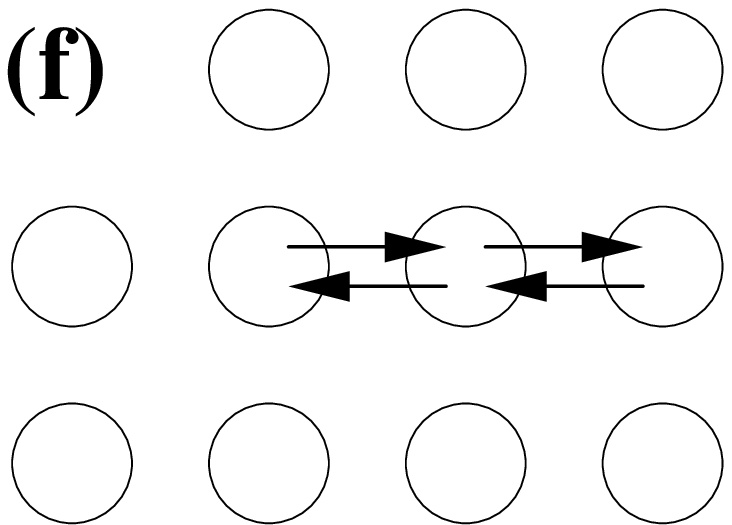}
\caption{\label{fig:EnD} Typical diagrams characterizing sets of tunneling
operations between neighboring sites in the two-dimensional square lattice
that appear in the energy correction of 2nd and 4th order. Tunneling operations
are denoted by arrows, lattice sites by circles. Diagrams contributing to the
perturbation expansion are those that can be interpreted as a single closed
path, {\it i.e}.\ all except (d).}
\end{figure}

The perturbation $\Vo$ to be considered consists of tunneling processes,
{\it i.e}.\ the annihilation of one particle at a given site in combination
with the creation of one particle at a neighboring site. Denoting a tunneling
operation by an arrow, one can visualize sets of tunneling operations graphically
by drawing diagrams. Obviously all process chains $P$ starting and ending at
the same (arbitrary) unperturbed state $|g\ra$ must contain the same number of
creation and annihilation processes at each site. Hence, those \emph{diagrams
contributing to the energy corrections contain closed paths only}. In
Fig.~\ref{fig:EnD} typical diagrams describing sets of tunneling operations in
the two-dimensional square lattice are sketched. The number of arrows
corresponds to the power of $J/U$ to which the diagram contributes. In
principle one can obtain all process chains $P$ appearing in the general
formula (\ref{eq:EnPro}) by generating all diagrams (sets of tunneling
operations) contributing to a given order in a first step, and by ordering the
operations of each diagram in all possible ways in a second step. Hence, before
energy corrections can be evaluated all contributing diagrams have to be
generated.

It has been noted that only diagrams containing closed paths of tunneling
operations contribute, since only these have the same number of creation and
annihilation operations at each site. In correspondence with the connected
cluster theorem \cite{GelfandEtAl90}, one can also show that \emph{only
connected diagrams give a non-vanishing contribution} to the energy correction,
{\it i.e}.\ those diagrams that cannot be divided into two or more sub-diagrams
with no lattice site in common or, in other words, that can be interpreted to
consist of a single closed path only. For example, the energy corrections
stemming from the different process chains that can be obtained from
diagram~(d) of Fig.~\ref{fig:EnD} must add up to zero. The basic idea for a
proof of this statement goes as follows \cite{GelfandEtAl90}: A connected
diagram would equally appear in the perturbation expansion for a system
differing from the one considered here by the fact that it consists of two
completely independent sub-systems that are not coupled to each other by 
tunneling and with one part of the diagram lying in each of them. In
that case, however, the perturbed state evolving from an unperturbed product
state will obviously be a product state with respect to both decoupled
sub-systems, and its energy will be the sum of both sub-system energies. Thus,
contributions to the energy depending in a non-additive way on the properties
of both sub-systems cannot occur. 

\begin{figure}[t]\centering
\includegraphics[width = 1\linewidth]{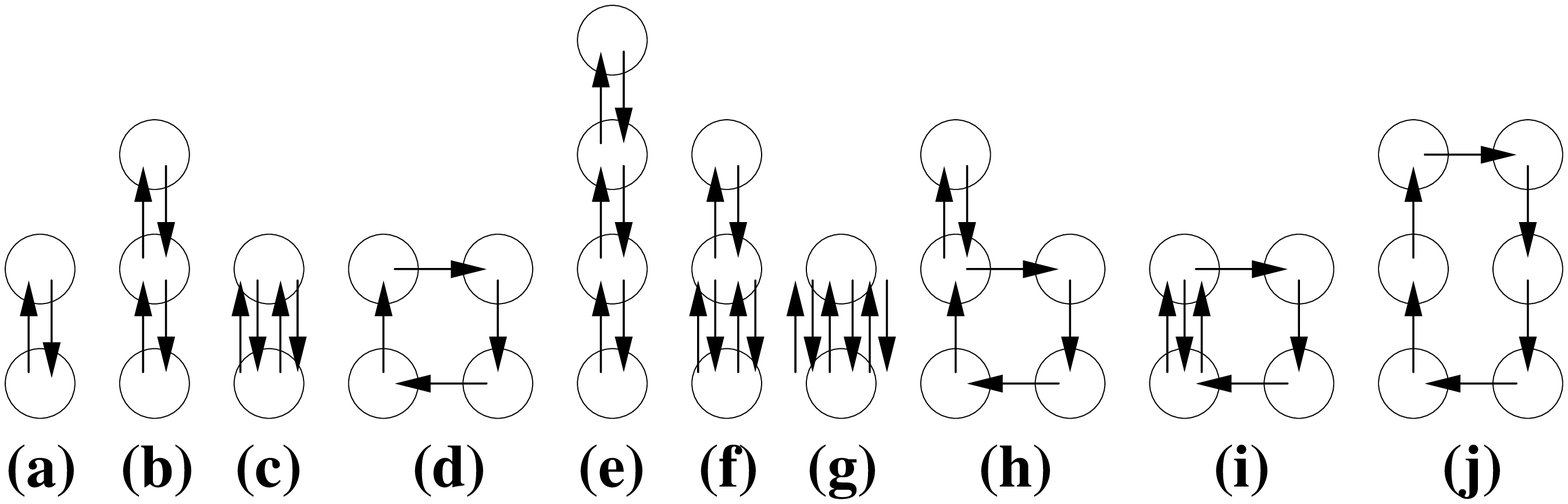}
\caption{\label{fig:EnTD} All topological diagrams contributing to the energy
correction of a hypercubic lattice in the leading orders 2, 4 and~6.} 
\end{figure}

Unless one is not dealing with a rather small system, computing
perturbative corrections to an extensive quantity like the energy will
generally involve too many diagrams to be accomplished in reasonable time. 
Nevertheless, it is possible to compute these corrections for a homogeneous
system with $\mu_i=\mu$ (or a system that shows a different kind of
translational symmetry). In that case topologically identical diagrams ---
like, {\it e.g}., (b) and (f) of Fig.~\ref{fig:EnD} --- will give identical
contributions. In Fig.~\ref{fig:EnTD} all types of topologically different
diagrams appearing in the leading orders\footnote{For any bipartite lattice
geometry, as the hypercubic one,
only an even number of tunneling operations leads back to the initial state such
that only energy corrections of even power in $J/U$ (not depending on the sign
of the tunneling matrix element) appear. This fact can also be exploited to
reduce the number of matrix elements considered in the general perturbation
expansion~(\ref{eq:EnPro}).} 2,~4, and~6 are plotted. The multiplicity $M_T$ of a
topological diagram $T$ is a weight factor being defined as the number of ways
it can be embedded into the given lattice geometry.
Note that disconnected diagrams would give rise to unphysical multiplicities
increasing with a power larger than one with the system size. 
In the following diagrams like those of Fig.~\ref{fig:EnD} containing operations
that are located in the lattice will be called \emph{geographical diagrams},
while diagrams like those of Fig.~\ref{fig:EnTD} that are characterized by
topology only will be called \emph{topological diagrams}.

\subsection{Computing high-order energy corrections}
In order to obtain all topological diagrams contributing to the energy
correction and their multiplicities on a computer, one needs two basic
routines, R2a and R2b, that will also serve to evaluate corrections to other
expectation values than that of the energy.
The first routine (R2a) computes all paths through the lattice via neighboring
sites starting from a given site $i$ to another site $j$ containing $\nu$
steps. By choosing $i=j$ and associating each step with a tunneling operation, in
that way one obtains all sets of tunneling operations, {\it i.e}.\ all
geographical diagrams, contributing in order $\nu$. However, two corrections
have to be taken into account. First, for closed loops a path visiting $s$
different lattice sites could equally be associated to each of the $s-1$ sites
different from $i$. Hence, such a diagram must be weighted by a factor of
$s^{-1}$. Second, it might happen that different paths contain exactly the
same tunneling operations (just in a different order). An example for that is
given by the diagram shown in Fig.~\ref{fig:EnD}~(b) that --- starting from the
site all arrows are connected to --- might be obtained by first moving
vertically and then horizontally or vice versa. If two paths include the same
tunneling operations, only one of them should be taken into account. For that
purpose a further routine (R2b) is needed that identifies paths containing the
same tunneling operations. 

For a homogeneous Bravais lattice all sites are equal and it suffices to 
consider just a single site $i$; moreover only topologically distinct diagrams
give distinct contributions. The relevant topological diagrams and their
multiplicities can be obtained by collecting geographical ones of the same
topology. Note that this step serves only to reduce the number of diagrams that
have to be evaluated. Hence, the algorithm used to probe the topological
equivalence of two geographical diagrams does not need to be perfect. A very
simple way of identifying identical topologies is to enumerate the sites
appearing in a geographical diagram by a single index in the order they appear
the first time in an associated path. Then routine R2b can be used to compare
the diagrams. For the closed loop diagrams contributing to the ground state
energy, with site $i$ not being distinguished from the others ones appearing in
the diagram, this approach has to be improved by probing enumerations
starting at different sites. 

One advantage of the diagram generation via paths consist in that fact that it
is easily implemented, even for high spatial dimensionalities $d$. Assuming,
for example, a hypercubic lattice, it is not difficult to design a routine R2a
that in principle works for arbitrary $d$ and also practically allows to
consider values of $d$ well above 3 that might be interesting to study the 
convergence towards meanfield behavior.

Once all diagrams of a given order have been obtained, their contribution to
the energy~(\ref{eq:EnPro}) can be evaluated by a last routine (R3). Each
sequence of the operations contained in a diagram corresponds to a different
process chain $P$, with the permutation of two identical operations --- as they
appear, {\it e.g}., in Fig.~\ref{fig:EnTD}~(c) --- not giving a new
process chain. Thus, the diagrams (a), (b), and (c) of Fig.~\ref{fig:EnTD} give
rise to $2!=2$, $4!=24$, and $4!/(2!)^2=6$ different process chains,
respectively. By applying the processes of a given chain one after the other to
a small array of occupation numbers $\{n_i\}$ initialized with $n^{(0)}_i=g_i$,
giving a sequence $\{n_i^{(0)}\},\{n_i^{(1)}\},\{n_i^{(2)}\},\ldots $, one can
compute (i) the matrix elements $V^{(m_k)}_{e_{k+1},e_k}$ [being
$-(J/U)\big(n^{(k)}_i(n^{(k)}_j+1)\big)^{1/2}$ for tunneling from site $i$ to
site $j$], (ii) the unperturbed energy differences $E_{e_k}-E_g$ of the
intermediate states [given by
$\sum_i \big(n_i^{(k)}(n_i^{(k)}-1)-g_i(g_i-1)-(\mu/U) (n_i^{(k)}-g_i)\big)$],
and (iii) whether an intermediate state $|e_k\ra$ equals $|g\ra$ ({\it i.e}.\
whether $n^{(k)}_i=g_i$ for all $i$) or not. Afterwards one can choose only
those matrix elements $(\alpha_{n-1},\ldots,\alpha_2,\alpha_{1})$ appearing in
the minimal expression~(\ref{eq:RedEn}) that have $\alpha_k=0$ if
$|e_k\ra=|g\ra$ and $\alpha_k\ne0$ if $|e_k\ra\ne|g\ra$ for all intermediate
states $k=1,2,\ldots,n-1$. Only these give a non-vanishing contribution to
the energy correction~(\ref{eq:EnPro}) that now can be evaluated with the help of
Eq.~(\ref{eq:EnergyWeights}) employing the energy differences $E_{e_k}-E_g$.

The scheme described in the preceding paragraph is not affected by the filling (the averaged particle number per
site), since no representation of a quantum-mechanical state space is needed.
Moreover, while applying the process chain to a set of occupation
numbers those unperturbed basis states $|\{n_i\}\ra$ that are relevant for a
given order of perturbation calculus (and only those) are generated
automatically.

\subsection{Corrections to expectation values and static susceptibilities}

Perturbative corrections to an expectation value $\la\Ao\ra$ (or a static
susceptibility $\chi_{\Ao,\Bo}$) in $\nu$th order of the perturbation
$\Vo$ can be obtained by computing energy corrections for the combined
perturbation $\Vo'=\lambda\Vo+x\Ao+y\Bo$ in first order in $x$ (and $y$) and in
$\nu$th order in $\lambda$, as expressed in Eq.~(\ref{eq:ExpA}) [and
Eq.~(\ref{eq:SusAB})] of section~\ref{sec:perturbation}. 
Hence, in order to compute such quantities, one can proceed exactly as before.
Now just one process $A_{e',e}|e'\ra\la e|$ associated to the
operator $\Ao\equiv\sum_{e,e'} A_{e',e}|e'\ra\la e|$ (and another one
associated to $\Bo$) has to be included in each process chain, in addition to
$\nu$ tunneling processes stemming from $\Vo$. The weight factors $G_{\{\alpha_k\}}^\text{min}$
entering the general perturbation expansion (\ref{eq:EnPro}) are those
referring to energy corrections of order $\nu+1$ for the computation of
expectation values (or $\nu+2$ for susceptibilities).

\begin{figure}[t]\centering
\includegraphics[width = 0.3\linewidth]{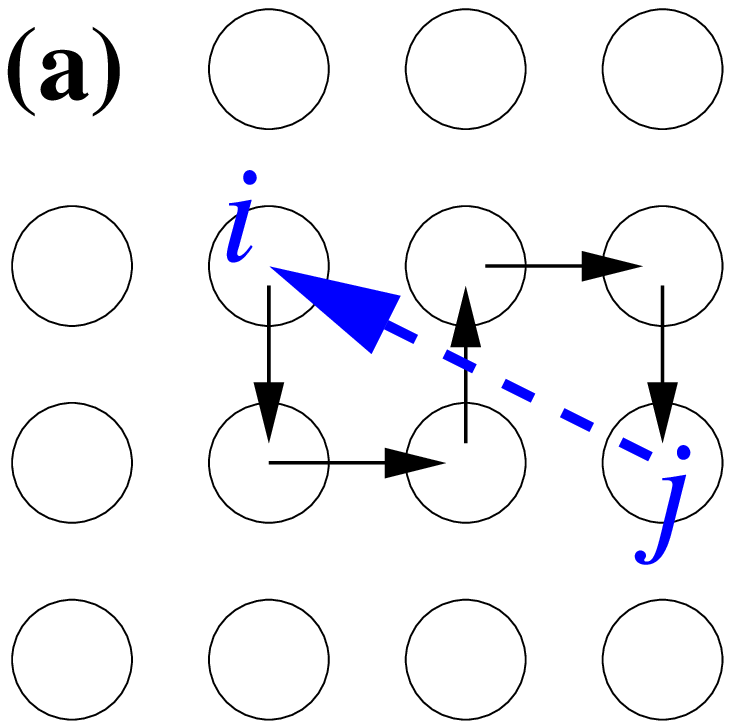}
\hspace{5ex}
\includegraphics[width = 0.3\linewidth]{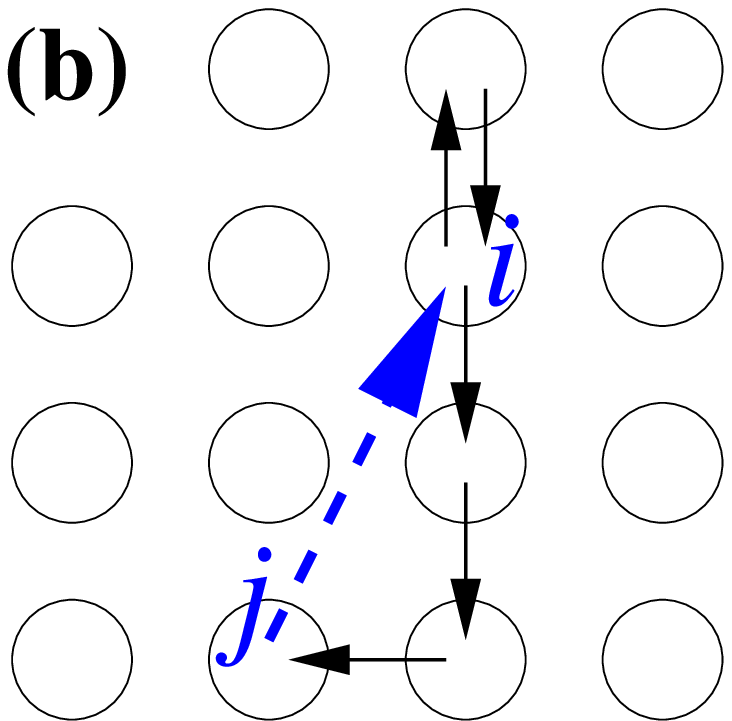}
\\\vspace{5ex}
\includegraphics[width = 0.3\linewidth]{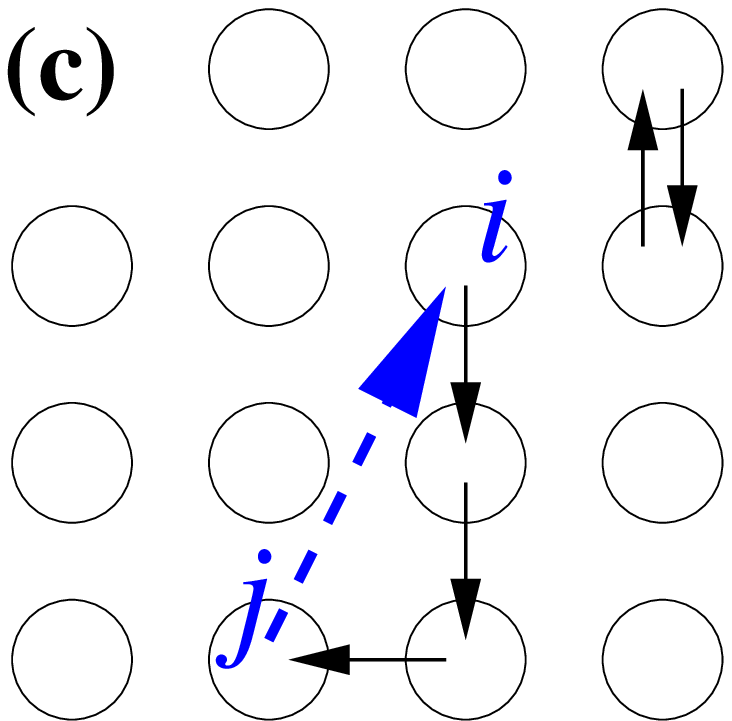}
\hspace{5ex}
\includegraphics[width = 0.3\linewidth]{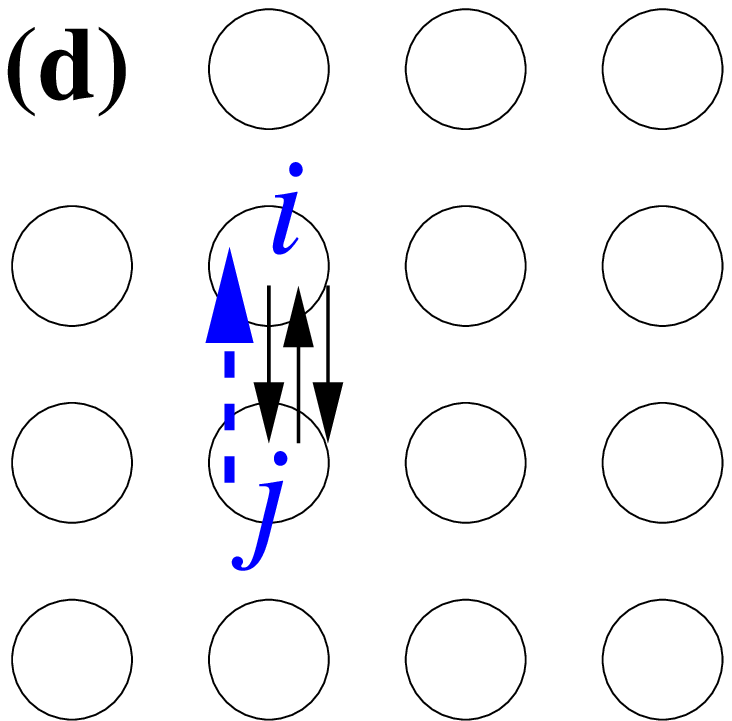}
\caption{\label{fig:KorrD} Typical diagrams appearing in the perturbation
expansion of correlation functions $\la\ba_i\hat{b}_j\ra$ in a 2D square
lattice. The dashed arrow is associated to an operator $\ba_i\hat{b}_j$. Again,
all diagrams contain closed loops only, with the solid arrows describing a path
from site $i$ to site $j$, and disconnected diagrams like (c) give
zero-contribution.}
\end{figure}

For the computation of a correlation function $\la\ba_i\bo_j\ra$, the
additional operation to be taken into account is the transfer of a particle from
site~$j$ to site~$i$ described by the operator $\ba_i\bo_j$. Denoting such an
operation by a dashed arrow, one can again use diagrams to describe sets of
operations. Typical diagrams are shown in Fig.~\ref{fig:KorrD}. As before, the
requirement that any sequence of the operations contained in a diagram must
lead from a given unperturbed state $|g\ra$ back to it ({\it i.e}.\ that the
number of particle annihilations equals that of particle creations at every
site) ensures that only closed loops (containing the dashed as well as solid
arrows) contribute. Since, moreover, again disconnected diagrams like (c) don't
need to be considered, the tunneling operations stemming from $\Vo$ can be
interpreted as a single path leading from $i$ to $j$. 
(One immediately sees that the perturbative treatment in the tunneling term
discussed here is not sensitive to single-particle correlations between sites
that are more than $\nu$ steps between neighboring sites apart.)
Also the generation of diagrams via the generation of paths through the lattice
can be accomplished in a similar way as before by using routine R2a. Paths lead
now from $i$ to $j$ and must not be corrected by the weight $s^{-1}$, since no
other starting point than the distinguished site $i$ will be taken into
account. Note that if $i$ and $j$ are neighboring sites, as it is the case in
diagram~(d) of Fig~\ref{fig:KorrD}, the operations related to the dashed arrow
still gives a factor of $x$ rather than $\lambda$, such that it is well
distinguished from a parallel tunneling operation stemming form $\Vo$ described
by a solid arrow. Therefore, {\it e.g}.\ in diagram (d) the permutation of both
upward tunneling operations (the ``solid'' and the ``dashed'' one) in a sequence
does lead to a new process chain, such that here $4!/2!=12$ different process
chains have to be taken into account.

\begin{figure}[t]\centering
\includegraphics[width = 0.3\linewidth]{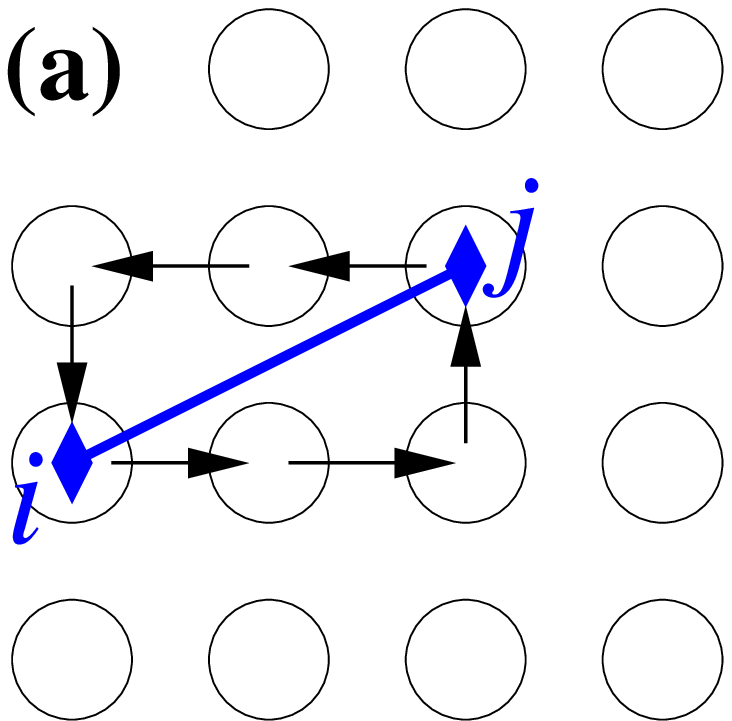}
\hspace{5ex}
\includegraphics[width = 0.3\linewidth]{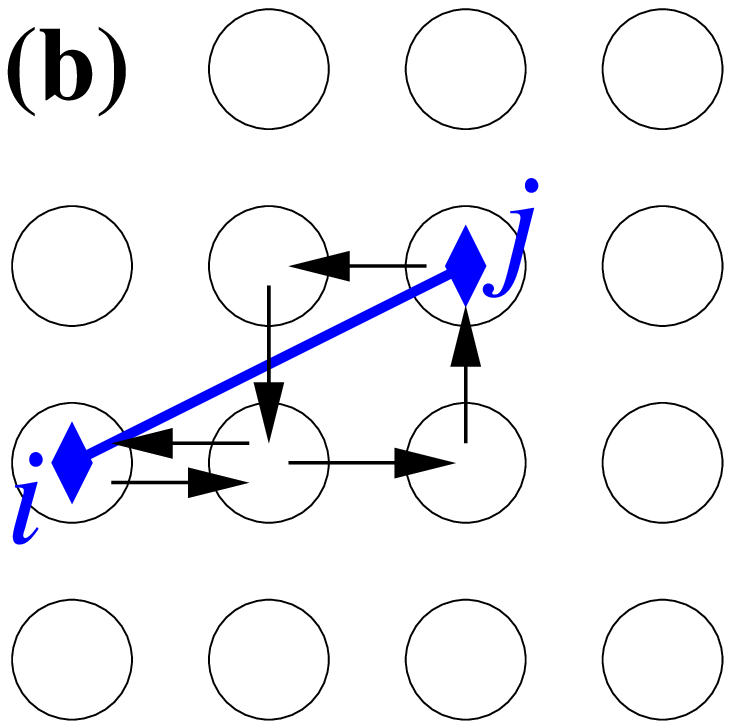}
\\\vspace{5ex}
\includegraphics[width = 0.3\linewidth]{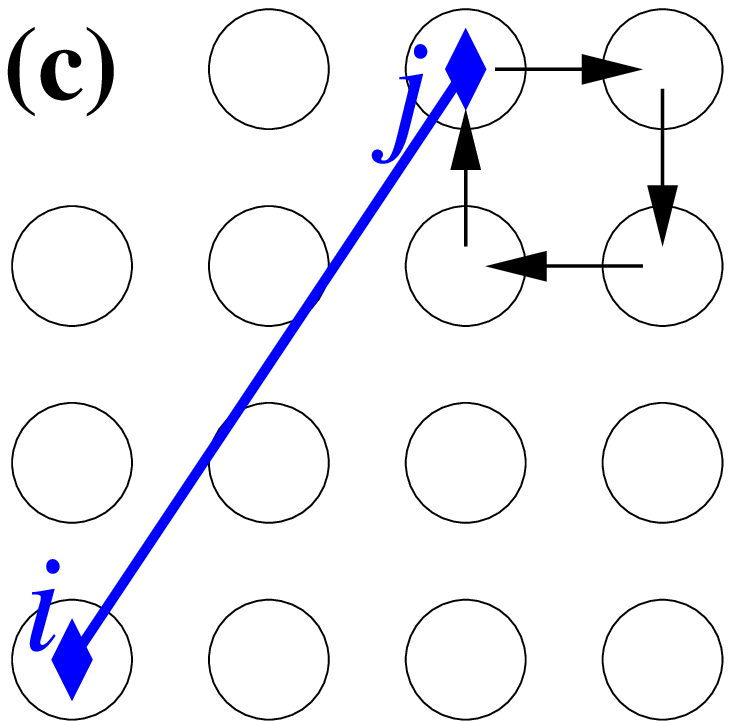}
\hspace{5ex}
\includegraphics[width = 0.3\linewidth]{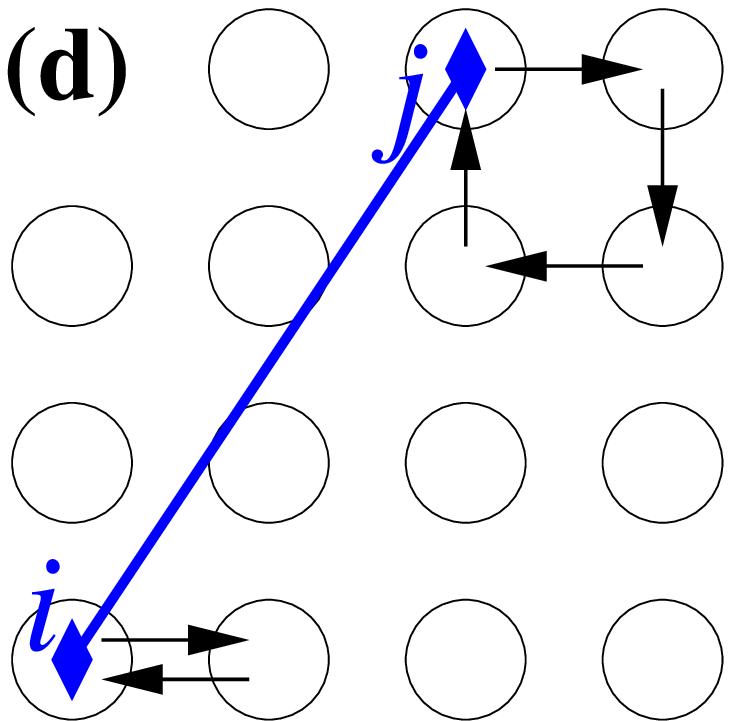}
\caption{\label{fig:NumD} Typical diagrams appearing in the perturbation
expansion of the number correlations $\la\no_i\no_j\ra$ in a 2D square
lattice. The operation associated with the linked diamonds is described by the
operator $\ba_i\hat{b}_i\ba_j\hat{b}_j$ that leaves occupation numbers
unaltered.}
\end{figure}

Another example is the computation of number correlations
$\la\no_i\no_j\ra$. The corresponding operator
$\no_i\no_j=\ba_i\bo_i\ba_j\bo_j$, with matrix elements depending on the
occupation of both sites $i$ and $j$, leaves any unperturbed occupation number
state unaltered. Nonetheless one can associate this ``operation'' with a
diagrammatic symbol that is chosen to be given by two diamonds at sites $i$ and
$j$, connected by a line. Fig.~\ref{fig:NumD} shows some diagrams appearing in
the perturbative expansion of expectation values $\la\no_i\no_j\ra$. Obviously,
the tunneling operations must form closed paths, such that the number of
creation and annihilation operations at each site are equal. Since, moreover,
again only connected diagrams contribute, the closed tunneling paths must
visit either $i$ or $j$. For given $i$ and $j$, one can generate all
contributing diagrams by generating all combinations of two paths, such that
one leads from $i$ back to $i$ and the other from $j$ back to~$j$ (including
paths of zero length). 

In the spirit of the examples treated so far, single-site expectation values
$\la\no_i^p\ra$ with arbitrary power $p$ are obtained from diagrams that are
generated by paths leading from site~$i$ back to site~$i$.

It is worth mentioning that the expectation value of an operator like
$\no_i\no_j$, acting on a few sites only, can be computed even in the case of a
large inhomogeneous system (provided, of course, perturbation theory is
meaningful). Since the contributing geographic diagrams are just those exploring
the neighborhood of $i$ and $j$, their number is limited and does not depend on
the system size. 
Correlations between $i$ and $j$ that are induced by the perturbing coupling
term (like $\la\ba_i\bo_j\ra$ in the present case) will, however, only be taken
into account in orders of perturbation theory that are comparable to the
distance between $i$ and $j$ (measured in steps between neighboring sites).
This directly reflects the limitation of the perturbative approach to systems
with such correlations decaying on a distance being at most equal to the order
of perturbation theory (at least as long as additional extrapolation techniques
are not applicable or considered).

\begin{figure}[t]\centering
\includegraphics[width = 0.3\linewidth]{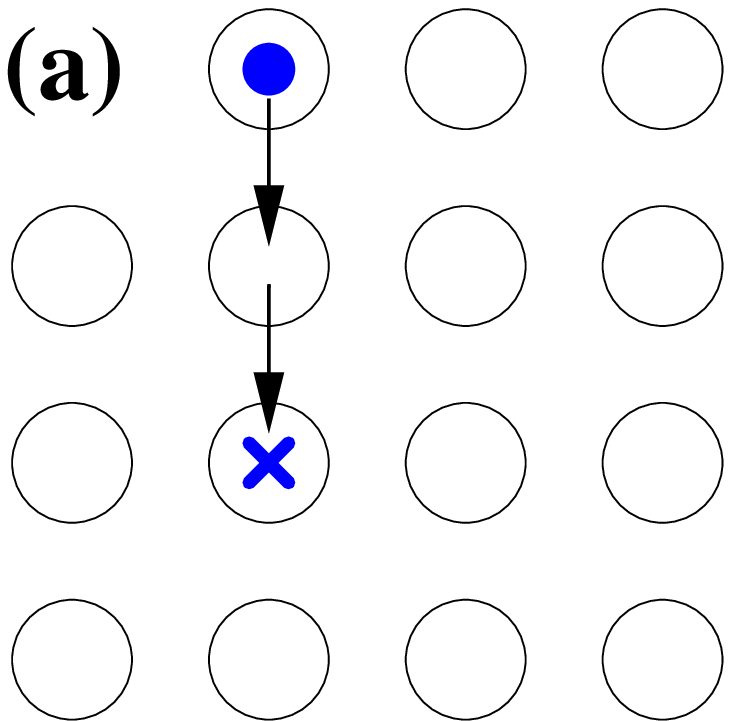}
\hspace{5ex}
\includegraphics[width = 0.3\linewidth]{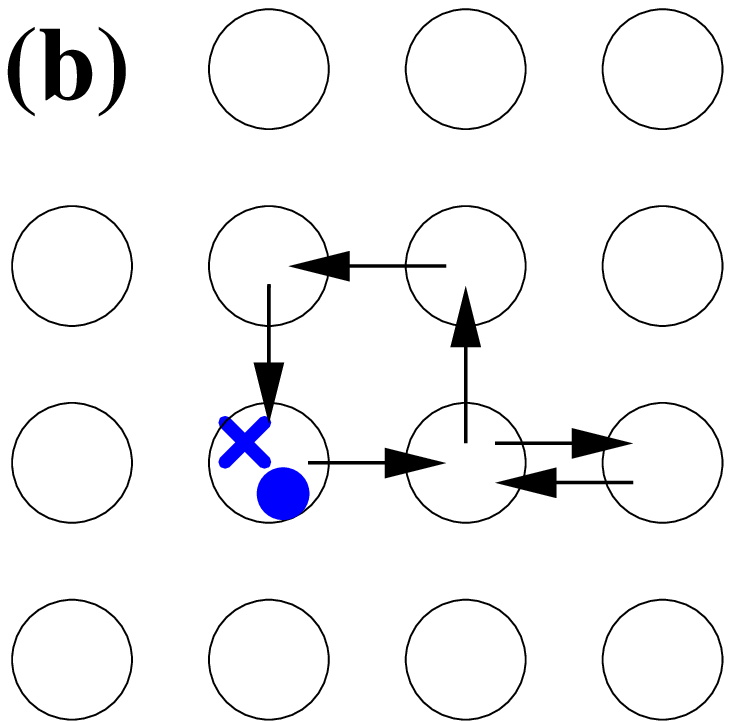}
\\\vspace{5ex}
\includegraphics[width = 0.3\linewidth]{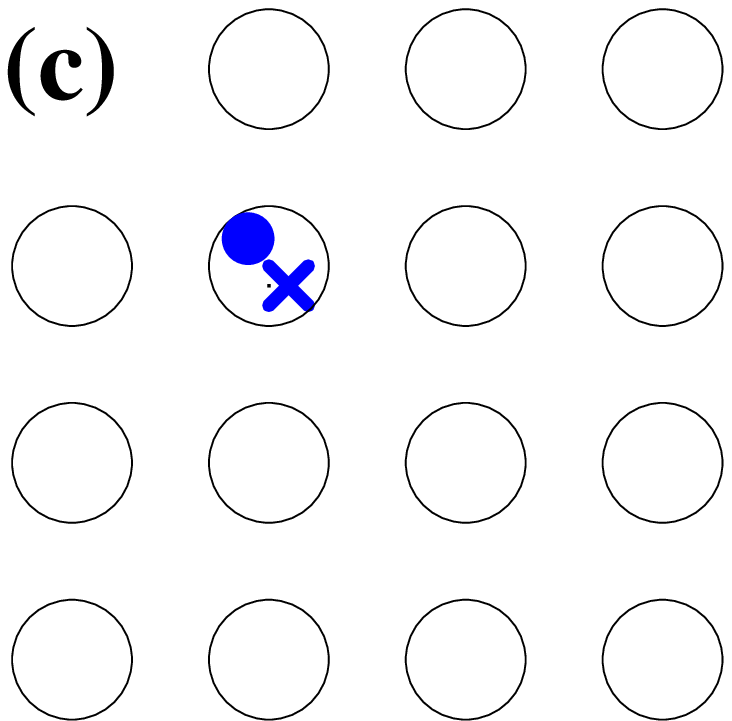}
\hspace{5ex}
\includegraphics[width = 0.3\linewidth]{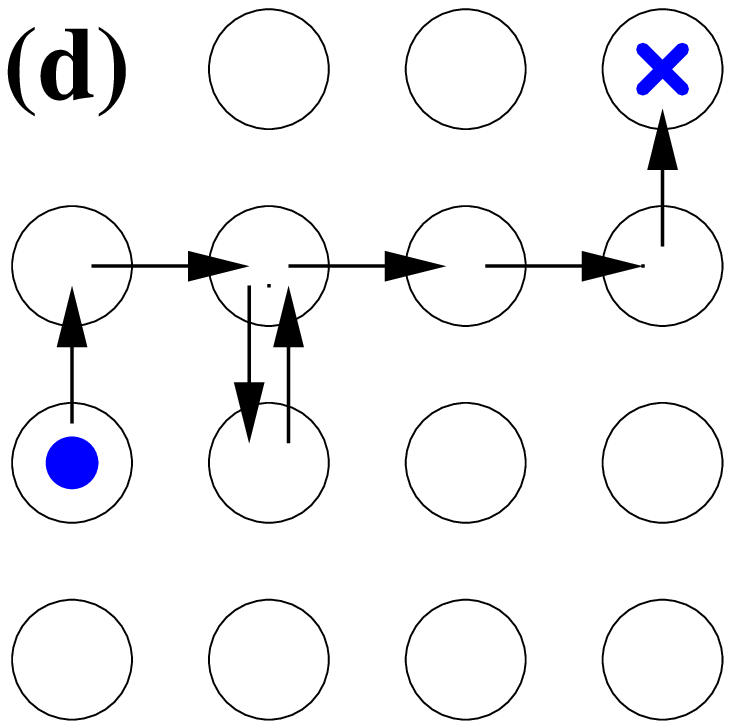}
\caption{\label{fig:PelD} Diagrams contributing to the static susceptibility
$\chi_{\hat{a}_0,\aa_0}$ that indicates the quantum phase transition from a
Mott-insulator to a superfluid.}
\end{figure}

\begin{figure}[t]\centering
\includegraphics[width = 0.8\linewidth]{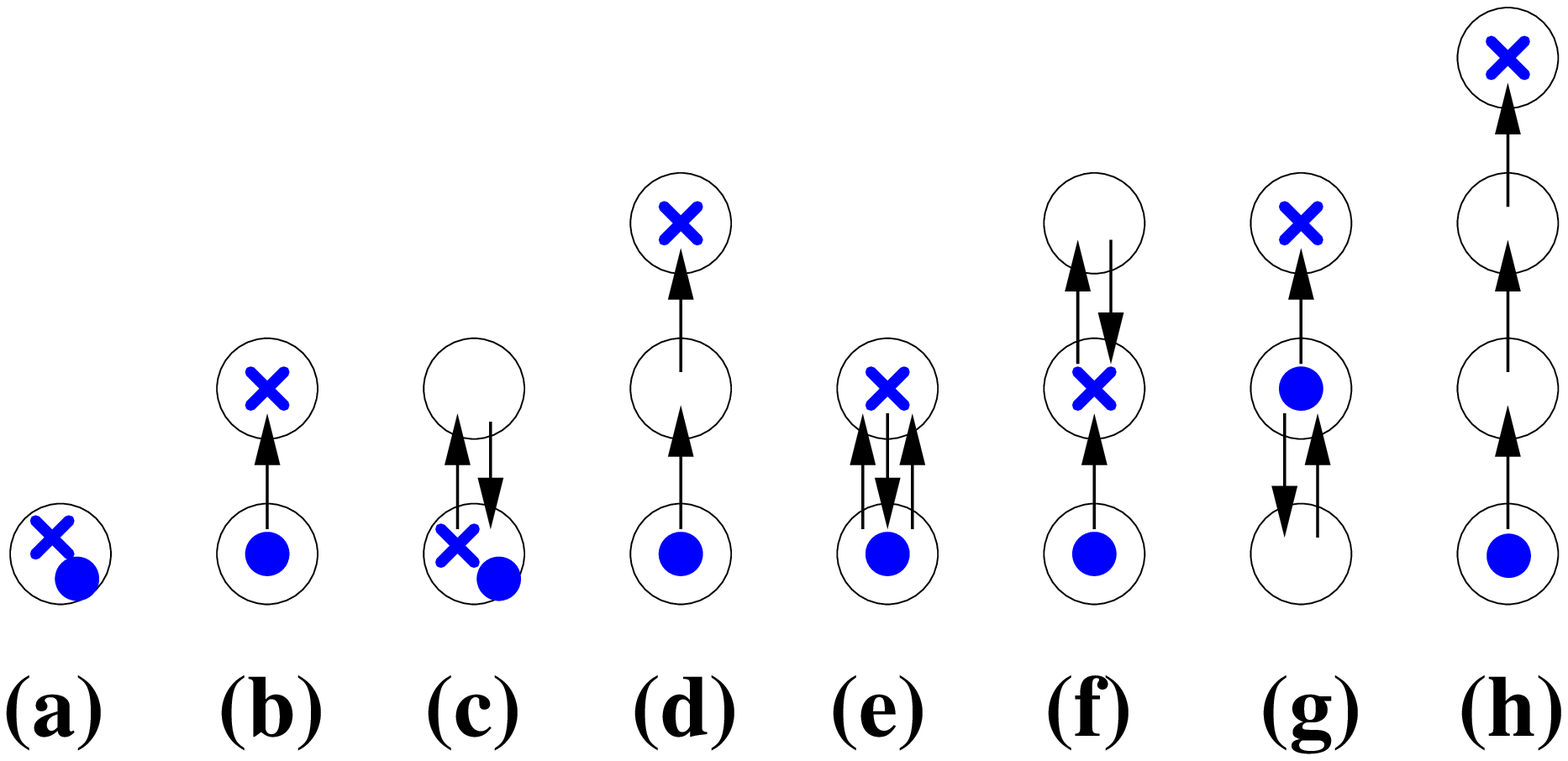}
\caption{\label{fig:PelDT} All topological diagrams contributing to the 
static susceptibility $\chi_{\hat{a}_0,\aa_0}$ in the leading orders 
0, 1, 2 and~3.} 
\end{figure}

Finally, it shall be outlined how the static susceptibility
$\chi_{\ao_0,\aa_0}\equiv \chi$ for the annihilation and creation operators
of the condensate mode, $\ao_0$ and $\aa_0$, with $\hat{a}_0\propto\sum_i\bo_i$ 
for the homogeneous system can be computed. This quantity is
proportional to the contribution  $\propto|\xi|^2$ to the energy obtained from
the effective Hamiltonian $\Ho_\text{eff}=\Ho+\sum_i(\xi\bo_i+\xi^*\ba_i)$. The
process chains contributing to it contain (apart from tunneling processes) one
creation process and one annihilation process that will be represented
diagramtically by a bullet $\bullet$ and a cross $\times$, respectively.
Examples for relevant diagrams are given in Fig.~\ref{fig:PelD}. They can be
obtained from connected paths starting and ending anywhere in the lattice
[including the zeroth order contribution shown in diagram (c)]. All
topological diagrams appearing in the leading orders 0, 1, 2, and 3 are shown
in Fig.~\ref{fig:PelDT}. In the limit of large spatial dimensionalities $d$,
tunneling several times along the same bond becomes very unlikely such that
diagrams like (a), (b), (d), and (h) of Fig.~\ref{fig:PelDT} that can be
interpreted as a path visiting each site only once give the major contribution
to the susceptibility.

For $d\ge2$ the susceptibility $\chi$ diverges when $J/U$ reaches some critical
value $(J/U)_\text{c}$, indicating the quantum phase transition from a
Mott-insulator to a superfluid \cite{NegeleOrland,dosSantosPelster09}.
However, the approximate value of $\chi$ in $\nu$th order,
$\chi^{(\nu)}\equiv\sum_{k=0}^\nu \beta_{k}\;(J/U)^{k}$ with the coefficients
$\beta_k$ depending on $\mu/U$, will always be finite for finite $J/U$ as long
as $\mu/U$ is non-integer. Thus, in order to extract the critical parameter
$(J/U)_\text{c}$ one has to resort to extrapolation to infinite order $\nu$.
The critical parameter can be associated with the radius of convergence of
the series for $\chi$ with respect to $J/U$, namely
$(J/U)_\text{c}=\lim_{k\to\infty}\beta_{k-1}/\beta_k$, assuming all
$\beta_k$ to have the same sign. Plotting $\beta_{k-1}/\beta_k$ versus $1/k$,
one finds to good approximation all data points to lie on a straight line,
suggesting the very simple phenomenological extrapolation scheme to extend the
line to $1/k=0$ by a linear fit\footnote{Since the slope of the line is
relatively low, the coefficients $\beta_k$ resemble approximately those of
a geometric series. Thus, also an exponential fit to the $\beta_k$ already gives
a good estimate to the phase boundary.}. This procedure gives the
phase boundary $(J/U)_\text{c}$ versus $\mu/U$ with an estimated error of 1-2
percent for arbitrary large filling $n$ and spatial dimensionalities $d=2$, 3,
and greater \cite{TeichmannEtAl09}. The errors have been estimated by monitoring
deviations of the approximate phase boundary while successively taking into
account more and more coefficients $\beta_{k}$. For $n=1$ these results agree
with those obtained by a strong coupling expansion \cite{ElstnerMonien99}
\footnote{Though using series expansions in powers of $J/U$, the approach of
Ref.~\onlinecite{ElstnerMonien99} is different from the one described here.
Instead of computing the susceptibility $\chi$ to obtain the phase boundary, the
authors compare the energy of a state with integer filling to the energies of
defect states having one additional particle or hole. Moreover, they use a
connected cluster expansion to obtain energy corrections.} 
($d=2$) and by Quantum Monte Carlo simulations
\cite{CapogrossoSansoneEtAl07,CapogrossoSansoneEtAl08} ($d=2$ and 3).
This simple example illustrates that extrapolation can be a valuable tool
augmenting high-order perturbation calculus. A brief introduction to more
advanced extrapolation techniques as well as further references can be found in
chapter 1 of Ref.~\onlinecite{OitmaaHamerZheng}.

\subsection{More general quantum lattice models\label{sec:models}}
So far, in this section the method has been developed in terms of the 
Bose-Hubbard model (\ref{eq:BH}). However, the approach is not restricted to
this model, and following along the lines of the above example, it can be
applied to a variety of quantum lattice models of the form given by
Eq.~(\ref{eq:Hgen}). 
This includes fermionic Hubbard models or Heisenberg type
spin models. 
In the remaining part of this section the properties of quantum
lattice models that are amenable to the method described in this paper will be
sketched. 

First of all the splitting of the full Hamiltonian (\ref{eq:Hgen}) into an
unperturbed part $\Ho_0$ and a perturbation $\Vo$ does not necessarily have to
be such that $\Ho_0$ contains just all on-site terms $\ho_i$, while $\Vo$
covers all coupling terms $\vo_{ij}$. Both terms $\Ho_0$ and $\Vo$ can
contribute to both on-site terms $\ho_i$ and coupling terms $\vo_{ij}$.
Moreover, the site index $i$ can be generalized to run over several degrees of 
freedom at every lattice point, or, similarly, the sum $\sum_{\la ij\ra}$ can
be extended to include not only pairs of nearest neighbors, but also further
pairs of near sites. It is, however, required that the relevant set of
eigenstates of $\Ho_0$ is characterized by a set of on-site quantum numbers
$\{n_i\}$ taking values $n_i=n_i^\text{min}$, $n_i^\text{min}+1$, 
$n_i^\text{min}+2$, \ldots, $n_i^\text{max}$ (with the possibility of arbitrary
large on-site state-space dimensionalities 
$\mathcal{D}_i=n_i^\text{max}-n_i^\text{min}+1$). In other words, $\Ho_0$ should
be expressed in terms of number operators $\no_i$ with
$\no_i|n_1n_2\cdots n_i \cdots\ra=n_i|n_1n_2\cdots n_i \cdots\ra$. For each site
$i$ there should also be a pair of ladder operators, $\hat{\ell}_i^{+}$ and
$\hat{\ell}_i^{-}$, being defined by
$\hat{\ell}_i^{\pm} |n_1n_2\cdots n_i \cdots\ra 
= \eta^{\pm}_i|n_1n_2\cdots n_i\pm1 \cdots\ra$. The perturbation $\Vo$
can be expressed in terms of both number and ladder operators. Given the
structure described in this paragraph, it will be possible to define diagrams
and to evaluate them in a similar fashion as described for the Bose-Hubbard
model above.

Lattice systems covered by the scheme just outlined are bosonic or fermionic
Hubbard models as well as spin Hamiltonians. In the former case the $n_i$ are
just occupation numbers runing from $n_i^\text{min}=0$ to
$n_i^\text{max}=\infty$ for bosons and $n_i^\text{max}=1$ for fermions. The
index $i$ can also distinguish between different internal degrees of freedom
(or species) of particles. For spins, $n_i$ would be associated with the
magnetic quantum number characterizing the spin at site $i$ along a
distinguished quantization axis, taking values between $\pm S$ with half-integer
total spin $S$. Concerning just the implementation, the systems (or the
unperturbed states) amenable to the approach described here do not need to be
homogeneous and they can be defined on various lattice geometries; also
frustration, disorder, and certain types of long-range interaction can be
present. But, of course, apart from being implementable the perturbation
expansion must as well be a suitable approximation scheme for a given problem.

For particles ({\it i.e}.\ in the Hubbard case) the unperturbed Hamiltonian
$\Ho_0$ can contain site-dependent potential terms $\propto\no_i$ and
two-particle density-density interaction terms $\propto\no_i\no_j$, also
three- and more particle terms are possible. In spin models, corresponding terms
can be considered, describing, {\it e.g.}, local magnetic fields along the
quantization axis or Ising type coupling. Non-local density-density interaction
terms appearing in the unperturbed Hamiltonian $\Ho_0$ can, in fact, be
long-ranged. In that case the unperturbed energies computed during the
evaluation of a given diagram will depend on the unerturbed quantum numbers
$n_i$ at sites not contained in that diagram (these will be unaltered by the
operations contained in the diagram). Considering a homogeneous unperturbed
state, this will cause only little extra computational effort, for an
inhomogeneous state the additional effort will just
grow as the number of sites within the range of interaction. Clearly, the
perturbative approach is limited to such strongly correlated phases that can be
explored by starting from an unperturbed product state. However, the treatment
of the bosonic Mott-transition \cite{ElstnerMonien99,TeichmannEtAl09} is an
example showing that even the boundaries of such phases in parameter space can
be obtained by applying suitable extrapolation schemes. One should also note
that (near) degeneracies between the unperturbed state considered and other
unperturbed states can spoil the results obtained by perturbation theory. 
As a simple example, this
can happen for varying on-site potentials (or magnetic fields) that cause at
some sites two different quantum numbers to lead to similar unperturbed
energies. As a remedy, in these cases one might consider to include degenerecy
breaking terms in $\Ho_0$ and to subtract them again in the perturbation $\Vo$,
cf. chapter 8 of Ref.~\onlinecite{OitmaaHamerZheng} and references therein.

The perturbation $\Vo$ can contain the ladder operators $\hat{\ell}_i^{\pm}$.
For the Hubbard models these correspond to the bosonic or fermionic creation and
annihilation operators, for a spin model they are given by the raising and
lowering operators for the given quantization axis at site $i$. For bosonic and
spin models the factors $\eta_i^{\pm}$ just depend on the local quantum number
$n_i$. In the case of the fermionic Hubbard model, the factors
$\eta^{\pm}_i$ accompanying the creation or annihilation of particles
take the values +1 and -1, depending on all occupation numbers $\{n_i\}$
(according to a given convention for the ordering of all sites $i$).
Taking care of these signs will cause additional effort. 
While the unperturbed Hamiltonians $\Ho_0$ can contain long-range coupling
terms, the coupling between different sites $i$ and $j$ appearing in the 
perturbation $\Vo$ should be rather short-ranged, since the number of diagrams
to be evaluated grows rapidly with the number of coupling terms contained in 
$\Vo$. 
If the coupling between different sites $i$ and $j$ is of the familiar
form $\hat{\ell}_i^{+}\hat{\ell}_j^{-}$, diagrams can, again, be generated
conveniently by finding paths through the lattice as described above for the
Bose-Hubbard model. This form is, however, quite typical, as it describes both
hopping of particles as well as spin-spin coupling in spin directions
transverse to the quantization axis.

\section{Summary\label{sec:summary}}

Let us briefly recapitulate the three basic steps of the approach described
above in sections \ref{sec:perturbation} and \ref{sec:lattice}.
The first task to be accomplished is to generate the leading order energy
corrections (\ref{eq:RedEn}) as they appear in standard
Rayleigh-Schr\"odinger perturbation theory such that in every order $\nu$ only
a minimum number of different matrix elements (\ref{eq:sh}), each characterized
by $\nu-1$ non-negative integers $\alpha_k$, has to be taken into account. This
can be done, {\it e.g.}, by starting from Kato's expression (\ref{eq:Kato}),
and merging matrix elements that (via a decomposition into elementary matrix
elements) are identified to give identical contributions. 
The number of relevant terms can further be reduced if a priori knowledge
is available about the vanishing of all matrix elements appearing in certain
orders $\nu$ with, {\it e.g}., $\nu$ being even, odd, or smaller than some
value~$\nu_0$. The obtained results also serve for the computation of
expectation values and static susceptibilities.

The contributing matrix elements are interpreted as sums over process chains in
a classical state space containing only the unperturbed states and not their
superpositions, {\it cf}.~Eq.~(\ref{eq:pro}). 
For the lattice problems considered, this formulation allows one to organize
the perturbation expansion in terms of simple connected diagrams, each
representing a collection of different operations. The second and the final
third step to be accomplished are the generation and the evaluation of these
diagrams. It has been shown that the generation of diagrams can be put down
to the generation of paths through the lattice, a rather simple task that can be
done on a computer even for large spatial dimensionalities~$d$. Finally, the
evaluation of diagrams is straightforward; one has to go through all possible
sequences of the operations contained in a given diagram and map them
according to Eq.~(\ref{eq:pro}) to the terms of the general perturbation
expansion (\ref{eq:RedEn}) obtained before. Since this procedure does not
require a representation of the quantum-mechanical state space on the
sub-lattice associated to a given diagram, it is not affected by large
dimensionalities of the on-site state spaces.

\section{Conclusion\label{sec:conclusion}}
A method to compute high-order series expansions for ground state properties of
quantum lattice models has been described that is based on
Rayleigh-Schr\"odinger perturbation calculus. The approach can be divided into
three basic steps that have to be accomplished on a computer; each of them can
be implemented with reasonable effort. Since the treatment of high spatial
dimensionalities as well as of large lattice-site state-space dimensionalities
is not connected to serious difficulties, the presented approach complements the
well-known connected cluster method \cite{Marland81,GelfandEtAl90}. Recently,
the method described here has been used to compute the phase diagram of the
Bose-Hubbard model on a $d$-dimensional hypercubic lattice, describing ultracold
bosonic atoms in optical lattices \cite{TeichmannEtAl09}. It allowed not only to
monitor in detail the convergence towards both the quantum-rotor limit of
high filling $n$ and the meanfield limit of large $d$, but also provided
experimentally relevant data for two and three dimensional systems at moderate
filling $n=2-10$. 
However, as outlined in section \ref{sec:models} a wide class of quantum lattice
models, including Heisenberg-type spin and Hubbard-type tight-binding models
are amenable to the approach described in this paper. These models can be
frustrated and inhomogeneous and they can contain disorder as well as long-range
interaction of the density-density or Ising type.  
Especially in view of the enormous interest in quantum lattice systems made of
ultracold atoms \cite{LewensteinEtAl07}, the ease of treating three-dimensional
systems can make the method a valueable tool for current research.

\section*{Acknowledgements}
The author is grateful to M.\ Lewenstein for kind hospitality at 
ICFO-Institut de Ci\`encies Fot\`oniques and appreciates a Feodor Lynen
research grant of the Alexander von Humboldt foundation as well as financial
support by the Spanish MEC (Grant FIS2008-00784 "TOQATA", ESF-EUROQUAM program
FIS2007-29996-E "FERMIX"). He thanks V.\ Steenhoff for providing the high-order
data shown in table \ref{tab:NumTerm} and N. Teichmann for carefully reading the
manuscript. The author also warmly thanks M.\ Holthaus for his continuous
support at the Universit\"at Oldenburg where part of the work has been done.

\bibliography{../../../BoseEinstein.bib}
\end{document}